\documentclass[prd,a4paper,aps,showpacs,twocolumn,floats,nofootinbib]{revtex4}
\usepackage{epsfig}
\usepackage{amssymb,amsmath,latexsym,mathrsfs}
\usepackage{graphicx}
\usepackage{varioref,xr-hyper}

\newcommand{\al}{\alpha}

\newcommand{\de}{\delta}

\newcommand{\ga}{\gamma}

\newcommand{\ka}{\kappa}

\newcommand{\om}{\omega}

\newcommand{\ra}{\rightarrow}

\newcommand{\be}{\begin{equation}}
\newcommand{\ee}{\end{equation}}
\newcommand{\bea}{\begin{eqnarray}}
\newcommand{\eea}{\end{eqnarray}}
\newcommand{\bean}{\begin{eqnarray*}}
\newcommand{\eean}{\end{eqnarray*}}
\newcommand{\dd}{\partial}

\def\lsim{\;\raise 0.4ex\hbox{$<$}\kern -0.8em\lower 0.62 ex\hbox{$\sim$}\;}
\def\gsim{\;\raise 0.4ex\hbox{$>$}\kern -0.7em\lower 0.62 ex\hbox{$\sim$}\;}
\begin{document}

\title{The dynamical Casimir effect in braneworlds}
\author{Ruth Durrer}
\email{ruth.durrer@physics.unige.ch}
\affiliation{D\'epartement de
Physique Th\'eorique, Universit\'e de
Gen\`eve, 24 quai Ernest Ansermet, CH--1211 Gen\`eve 4, Switzerland}
\author{Marcus Ruser}
\email{marcus.ruser@physics.unige.ch}
\affiliation{D\'epartement
de Physique Th\'eorique, Universit\'e de
  Gen\`eve, 24 quai Ernest Ansermet, CH--1211 Gen\`eve 4, Switzerland}


\begin{abstract}
In braneworld cosmology the expanding Universe is realized as a 
brane moving through a warped higher-dimensional spacetime. 
Like a moving mirror causes the creation of photons out of 
vacuum fluctuations, a moving brane leads to graviton production. 
We show that, very generically, KK-particles scale like stiff matter
with the expansion of the Universe and can therefore not represent the
dark matter in a warped braneworld. 
We present results for the production of massless and Kaluza-Klein
(KK) gravitons for bouncing branes in five-dimensional Anti de Sitter
space. 
We find that for a realistic bounce the back reaction from the 
generated gravitons will be most likely relevant.
This Letter summarizes the main results and conclusions from numerical
simulations which are presented in detail in a long paper
[M.~Ruser and R.~Durrer, Phys. Rev. D {\bf 76}, 104014 (2007), arXiv:0704.0790].
\end{abstract}
%
\maketitle
%
%
{\bf Introduction:} 
String theory, the most serious candidate for a quantum theory of
gravity, predicts the existence of 'branes', i.e. hypersurfaces in the
10- (or 11-) dimensional spacetime on which ordinary matter, e.g. gauge
particles and fermions, are confined. Gravitons can move freely in the
'bulk', the full higher dimensional spacetime~\cite{Polchi}.
The scenario, where our Universe moves through a five-dimensional Anti de
Sitter (AdS)  spacetime has been especially successful in reproducing
the observed four-dimensional behavior of gravity. It has been shown
that at sufficiently low energies and large scales, not only gravity
on the brane looks four dimensional~\cite{RSI+II}, but also
cosmological expansion 
can be reproduced~\cite{bine}. We shall concentrate here on this
example and comment on behavior which may survive in
other warped braneworlds.
We consider the following situation: A fixed 'static brane' is
sitting in the bulk. The 'physical brane', our Universe, is first moving
away from the AdS Cauchy horizon, approaching the second brane. This
motion corresponds to a contracting Universe. After a closest
encounter the physical brane turns around and moves away from the
static brane.
This motion mimics the observed expanding Universe.
\\
The moving brane acts as a time-dependent 
boundary for the 5D bulk leading to production of gravitons from 
vacuum fluctuations in the same way a moving mirror causes photon 
creation from vacuum in dynamical cavities~\cite{casi}.  
Apart from massless gravitons, braneworlds allow for a tower of
Kaluza-Klein (KK) gravitons which appear as massive particles on the
brane leading possibly to phenomenological consequences.
\\
We postulate, that high energy stringy physics will lead to 
a turnaround of the brane motion, i.e., provoke a
repulsion of the physical brane from the static one. 
This motion is modeled by a kink where the brane velocity 
changes sign. As we shall see, a perfect kink leads to 
divergent particle production due to its infinite acceleration. 
We therefore assume that the kink is rounded off at the 
string scale $L_s$. Then particles with energies 
$E>E_s =1/L_s$ are not generated.
This setup represents a regular 'bouncing Universe' as, 
for example the 'ekpyrotic Universe'~\cite{ekpy}. 
Four-dimensional bouncing Universes have also been 
studied in Ref.~\cite{filo}.
\\
%
{\bf Moving brane in AdS$_5$:} 
Our starting point is the metric of AdS$_5$ in 
Poincar\'e coordinates:
\begin{equation}\label{e:bulk-metric}
 d s^2
 = g_{{AB}} d x^{{A}} d  x^{{B}}
 = \frac{L^2}{y^2} \left[-d t^2 + \delta_{ij} d x^i d x^j + d
 y^2\right]~.
\end{equation}
The physical brane (our Universe) is located at some time-dependent
position $y=y_b(t)$, while the static brane is at a 
fixed position $y=y_s > y_b(t)$.
The scale factor on the brane is
\begin{equation}
a(\eta)= \frac{L}{y_b(t)}\,, \;\,  d\eta =\sqrt{1-v^2}dt =
\ga^{-1}dt~,\;\, v =\frac{dy_b}{dt}~,
\nonumber
\end{equation}
where we have introduced the brane velocity $v$ and the conformal
time $\eta$ on the brane.
If $v\ll 1$, the junction
conditions lead to the Friedmann equations on the brane.
For reviews see~\cite{roy,myrev}. Defining the string
and Planck scales by $\ka_5 \equiv L_s^3$ and $\ka_4 \equiv L_{Pl}^2$
the Randall-Sundrum (RS) fine tuning condition~\cite{RSI+II} implies
\be\label{LLsLPl}
\frac{L}{L_s} = \left(\frac{L_s}{L_{Pl}}\right)^2~.
\ee
We assume that the brane energy density is dominated 
by a radiation component. The contracting ($t<0$) and 
expanding ($t>0$) phases are then described by
\begin{eqnarray}
a(t) &=& \frac{|t| + t_b}{L}\,, \qquad y_b(t) =
\frac{L^2}{|t| + t_b}\,,\label{e:yb}\\
v(t) &=& -\frac{\mathrm{sign}(t)L^2}{(|t|+t_b)^2} \simeq -H L~
\label{e:vb}
\end{eqnarray}
where $H=(da/d\eta)/a^2$ is the Hubble parameter and we have used that
$\eta\simeq t$ if $v \ll 1$. A small velocity also requires $y_b(t) \ll L$.
The transition from contraction to expansion 
is approximated by a kink at $t=0$, such that at 
the moment of the bounce
\begin{eqnarray}
|v(0)|\equiv v_b = \frac{L^2}{t_b^2}\,,\;\, 
a_b = a(0) = \frac{1}{\sqrt{v_b}} \,,\;\,
H_b^2 = \frac{v_b^2}{L^2}~.  \label{Hb}
\end{eqnarray}
\\
%
{\bf Tensor perturbations:}
We now consider tensor perturbations $h_{ij}$ on this background,
\begin{eqnarray}
 d s^2 = \frac{L^2}{y^2}
 \left[-d t^2+(\delta_{ij}+2h_{ij})d x^i d x^j+d y^2 \right]~.
\end{eqnarray}
For each polarization, their amplitude $h$ satisfies 
the Klein-Gordon equation in AdS$_5$~\cite{myrev}
\begin{equation}
 \left[\dd_t^2 +k^2 -\dd_y^2 + \frac{3}{y}\dd_y \right]h(t,y;{\bf k}) = 0~
\label{e:T-bulk-eq}
\end{equation}
where $k=|{\bf k}|$ is the momentum parallel to the brane 
and $h$ is subject to the boundary (2nd junction) conditions 
\begin{equation}
\left(v\partial_t + \partial_y\right)h |_{y_b(t)} = 0\;
\rightarrow \; \partial_y h |_{y_b(t)} = 0\;\;
{\rm and}\;\;
\partial_yh|_{y_s}=0~.
\label{e:2nd junction} 
\end{equation}
Being interested in late-time (low energy) effects, we have approximated 
the first of those conditions by a Neumann condition ($v\ll 1$).
Then, the spatial part of Eq.~(\ref{e:T-bulk-eq}) together with 
(\ref{e:2nd junction}) forms a 
Strum-Liouville problem at any given time and therefore has a 
complete orthonormal set of eigenfunctions
$\{\phi_\alpha(t,y)\}_{\alpha = 0}^\infty$. 
These 'instantaneous' mode functions are given by
\bea
\phi_0(t) &=& \frac{y_s y_b(t)}{\sqrt{y_s^2 - y_b^2(t)}}.
\label{zero mode phi}\\
\phi_n(t,y) &=& N_n (t) y^2C_2(m_n(t),y_b(t),y) ~ \mbox{ with}
\nonumber \\   \hspace*{-2mm}
C_\nu(m,x,y) &=& Y_1(m x) J_\nu(my)\! -\! J_1(m x) Y_\nu(m y)\,
\eea
and satisfy $[ -\partial_y^2 + (3/y)\partial_y] \phi_\alpha(y)
=m_\alpha^2\phi_\alpha(y)$ as well as (\ref{e:2nd junction}).
$N_n$ is a time-dependent normalization condition. 
More details can be found in~\cite{long}.
The massless mode $\phi_0$ represents the ordinary
four-dimensional graviton on the brane, 
while the massive modes are KK gravitons. 
Their masses are quantized by the
boundary condition at the static brane which 
requires $C_1(m_n,y_b,y_s) =0$.
At late times and for large $n$ the KK masses are roughly
given by $m_n  \simeq  n\pi/y_s$.
The gravity wave amplitude $h$ may now be decomposed as \cite{long}
\begin{equation}
h (t,y;{\bf k}) = \sqrt{\frac{\kappa_5}{L^3}}\sum_{\alpha = 0}^\infty
q_{\alpha,{\bf k}}(t)\phi_\alpha(t,y)
\label{e: mode expansion}
\end{equation}
where the prefactor assures that the variables $q_{\alpha,{\bf k}}$
are canonically normalized.
Their time evolution is determined by the brane 
motion [cf.~Eq.~(\ref{deq for q})].
\\
%
{\bf Localization of gravity:}
From the above expressions and using $L/y_b(t) =a(t)$, we can
determine the late-time behavior of the mode functions
$\phi_\alpha$ on the brane ($y_b\ll L\ll y_s$)
\bea
\phi_0(t,y_b) \ra \frac{L}{a} \;,\;\;
\phi_n(t,y_b) \ra \frac{L^2}{a^2} \sqrt{\frac{\pi m_n}{2y_s}}~.
\eea
At this point we can already make two crucial observations:  
First, the mass $m_n$ is a comoving mass. The instantaneous
energy of a KK graviton is $\om_{n,k} =\sqrt{k^2+m_n^2}$, where $k$
denotes comoving wave number. The 'physical mass' of a KK mode measured
by an observer on the brane with cosmic time $d\tau =adt$ is therefore
$m_n/a$, i.e. the KK masses are redshifted with the expansion of the
Universe. This comes from the fact that $m_n$ is the wave number
corresponding to the $y$ direction with respect to the bulk time $t$
which corresponds to {\it conformal time} $\eta$ on the brane and
not to physical time. It implies that the energy of KK particles on
a moving AdS brane is redshifted like that of massless
particles. 
From this alone we would expect the energy density of 
KK modes on the brane decays like $1/a^4$.

But this is not all. In contrast to the zero mode which behaves as 
$\phi_0(t,y_b) \propto 1/a$ the KK-mode functions 
$\phi_n(t,y_b)$ decay as $1/a^2$ with the expansion of 
the Universe and scale like $1/\sqrt{y_s}$.  
Consequently the amplitude of the KK modes on the brane 
dilutes rapidly with the expansion of the Universe and 
is in general smaller the larger $y_s$.
This can be understood by studying the probability of finding a 
KK-graviton at position $y$ in the bulk which turns out
to be much larger in regions of less warping
than in the vicinity of the physical
brane\cite{long}.
If KK gravitons are present on the brane, they escape 
rapidly into the bulk, i.e., the moving brane looses them, 
since their wave function is repulsed away from the brane.
This causes the additional $1/a$-dependence of 
$\phi_n(t,y_b)$ compared to $\phi_0(t,y_b)$.
The $1/\sqrt{y_s}$-dependence expresses 
the fact that the larger the bulk the smaller 
the probability to find a KK-graviton 
at the position of the moving brane.
This behavior reflects the localization of gravity: 
traces of the five-dimensional nature of gravity 
like KK gravitons become less and less 'visible' 
on the brane as time evolves. 
As a consequence, the energy density of 
KK gravitons at late times on the brane behaves as
\be
\rho_{\rm KK} \propto 1/a^6 ~.
\label{e:KK energy density}
\ee
It means that KK gravitons redshift like stiff matter and cannot
be the dark matter in an AdS braneworld since their energy density does 
not have the required $1/a^3$ behavior. They also do not behave like
dark radiation~\cite{roy,myrev} as one might naively expect. 
This new result is derived in detail in Ref.~\cite{long}.
It is based on the calculation of 
$\langle \dot{h}^2(t,y_b,{\bf k})\rangle \propto \langle 
\dot{q}^2_{\alpha,{\bf k}}(t)\rangle\phi^2_\alpha(t,y_b)$ 
where the bracket incorporates a quantum expectation value
with respect to a well-defined initial vacuum state and 
averaging over several oscillations of the field~\cite{long}. 
An overdot denotes the derivative with respect to $t$.
The scaling behaviour (\ref{e:KK energy density}) is due to
$\phi^2_\alpha(t,y_b)$ only, since graviton 
production from vacuum fluctuations has ceased at late times
(like in radiation domination) which is necessary for a 
meaningful particle definition. 
Then, $\langle \dot{q}^2_{\alpha,{\bf k}}(t)\rangle$ is related
to the number of produced gravitons and is constant in time.
In case that amplification of tensor perturbation is still
ongoing, e.g., during a de Sitter phase, the energy density related
to the massive modes  might scale differently.
The scaling behavior (\ref{e:KK energy density}) remains valid 
also when the fixed brane is sent off to infinity and we 
end up with a single braneworld in AdS$_5$,
like in the Randall-Sundrum~II scenario~\cite{RSI+II}. 
The situation is not altered if we replace the graviton by a scalar or
vector degree of freedom in the bulk. Since every bulk degree of freedom
must satisfy the five-dimensional Klein-Gordon equation, the mode
functions will always be the functions $\phi_\al$, and
the energy density of the KK-modes decays like $1/a^6$.
KK particles on a brane moving through an AdS bulk cannot play the role of dark
matter. 
\\
It is important here that we consider a static bulk and the
time depencence of the brane comes solely from its motion through the
bulk. 
In Ref.~\cite{lang} the situation of a {\em fixed brane} in a 
{\em time-dependent bulk} is discussed.
There it is shown that under certain assumption (separability 
of the $y$ and $t$ dependence of fluctuations), the energy 
density of KK modes on a low energy cosmological brane does scale 
like $1/a^3$ which seems to be in contradiction with our result.
However, the approximations used in \cite{lang} lead to a system 
of equations governing the expansion of the Universe but
neglecting the time dependence of the bulk.
The situation is then effectively four dimensional even for 
the KK modes; effects of the fifth dimension like the possibility 
of KK gravitons escaping into the bulk seem to be 
lost in this approach. 
In our case we would have a similar situation if we 
keep the expansion on the brane $a(t)$ but take the 
position of the brane in the bulk as static 
$y_b(t) = {\rm const}$,
which is not consistent with the general relation 
$y_b(t) = L/a(t)$.
[For a fixed physical mass $M=m/a$, if we
neglect the time dependence of $\phi_n(y_b(t)) \propto 1/a^2$
we also obtain an energy density for this
mass  proportional to $1/a^3$.]
\\
{\bf Particle production:}
The equation of motion for the canonical variables 
$q_{\alpha,{\bf k}}$ is of the form, see Ref.~\cite{long},
\begin{eqnarray}
\ddot{q}_{\alpha,{\bf k}} + \omega_{\alpha, k}^2 q_{\alpha,{\bf k}}
=  \sum_{\beta\neq\al} {\cal M}_{\al\beta}\dot{q}_{\beta,{\bf k}}
+\sum_\beta{\cal N}_{\alpha\beta}q_{\beta,{\bf k}} ~.
\label{deq for q}
\end{eqnarray}
Here $\om_{\al,k} = \sqrt{ k^2 +m_\al^2}$ is the frequency
of the mode and ${\cal M}$ and ${\cal N}$ are coupling
matrices.
When we quantize these variables, gravitons can be created by
two effects: First, the time dependence of the effective frequency
$(\om_{\al,k}^{\rm eff})^2 =\om_{\al,k}^2 -{\cal N}_{\alpha\al}$
and second, the time dependence of the mode couplings 
described by the antisymmetric matrix $ {\cal M}$ and the 
off-diagonal part of ${\cal N}$. 

Note that Equation (\ref{deq for q}) is derived from the 
corresponding action for the variables $q_{\alpha,{\bf k}}$  
rather than from the wave equation (\ref{e:T-bulk-eq}) itself.
In this way the approximated boundary conditions 
(\ref{e:2nd junction}) can be implemented consistently
\cite{long,cyril}.
\\
In the technical paper \cite{long} we have studied graviton 
production provoked by a brane moving according 
to (\ref{e:yb}) in great detail numerically.
We have found that for long wavelengths, $kL\ll 1$, 
the zero mode is mainly generated by its self-coupling, 
i.e. the time dependence of its effective
frequency. 
One actually finds that ${\cal N}_{00} \propto \de(t)$, so
that there is an instability at the moment of the kink 
which leads to particle creation, and the number of 
4D-gravitons is given by $2v_b/(kL)^2$. 
This is specific to radiation dominated expansion where 
$H^2a^2 = -\dd_\eta(Ha)$. For another expansion law we
would also obtain particle creation during the 
contraction and expansion phases.
Light KK gravitons are produced mainly via their
coupling to the zero mode.
This behavior changes drastically for short wavelengths 
$kL\gg 1$. 
Then the evolution of the zero mode couples strongly 
to the KK modes and production of 4D gravitons via the decay of 
KK modes takes place. In this case the number of 
produced 4D gravitons decays only like $\propto 1/(kL)$.
\\
%
{\bf Results and discussion:} 
The numerical simulations have revealed a multitude of
interesting effects.  
In the following we summarize the main 
findings.
We refer the interested reader 
to Ref.~\cite{long} for an extensive discussion.
\\
For the zero-mode power spectrum we find on 
scales $kL \ll 1$ on which we observe cosmological 
fluctuations (Mpc or larger) 
\begin{equation}
{\cal P}_0(k) =  \frac{\kappa_4}{2\pi^3} v_b\left\{\begin{array}{ll}k^2
& \mbox{ if }~ kt\ll 1 \\
\frac{1}{2}(La)^{-2} & \mbox{ if }~ kt\gg 1~. \end{array} \right.
\end{equation} 
The spectrum of tensor perturbations is blue 
on super-horizon scales as one would expect for an ekpyrotic 
scenario. On cosmic microwave background scales the 
amplitude of perturbations is
of the order of $(H_0/m_{\rm Pl})^2$ and hence unobservably small. 
\\
Calculating the energy density of the produced
massless gravitons one obtains~\cite{long}
\begin{equation}\label{4.15}
\rho_{h0} \simeq \frac{\pi}{2a^4}\frac{v_b}{LL_s^3}~.
\end{equation}
Comparing this with the radiation energy density,
$\rho_{\rm rad} = (3/(\kappa_4 L^2))a^{-4}$, the RS fine-tuning condition
leads to the simple relation
\begin{equation}
\rho_{h0} / \rho_\mathrm{rad}\simeq v_b/2.
\end{equation}
The nucleosynthesis bound~\cite{michele} requests $\rho_{h0} \lsim 0.1
\rho_\mathrm{rad}$, which implies $v_b\le 0.2$, justifying 
our low energy approach. 
The model is not severely constrained by the zero-mode. 
\\
More stringent bounds come from the KK modes. 
Their energy density on the brane is found to be
\begin{equation}
\rho_{\rm KK} \simeq \frac{\pi^5}{a^6} \frac{v_b^2}{y_s}\frac{L^2}{L_s^5}.
\end{equation}
This result is dominated by high energy KK gravitons 
which are produced due to the kink.   
It is reasonable to require that the KK-energy density on the brane 
be (much) smaller than the radiation density 
at all times, and in particular,
right after the bounce where $\rho_{\rm KK}$ is greatest.
If this is not satisfied, back reaction cannot be neglected.  
We obtain with $\rho_{\rm rad}(0) = 3H_b^2/\ka_4$
\begin{equation}
\left.\left( \frac{\rho_{\rm KK}}{\rho_{\rm rad}} \right)
\right|_{a=a(0)=1/\sqrt{v_b}} 
\simeq 100 \,v_b^3 \left(\frac{L}{y_s}\right)\left(\frac{L}{L_s}\right)^2 .
\label{e:constraint0}
\end{equation} 
If we use the largest value for the brane velocity $v_b$ 
admitted by the nucleosynthesis bound $v_b \simeq 0.2$ and require 
that $\rho_{\rm KK}/\rho_{\rm rad}$ be (much) smaller than one 
for back-reaction effects to be negligible, we obtain the 
very stringent condition
\begin{equation}
\frac{L}{y_s} < \left(\frac{L_s}{L}\right)^2.
\label{e:constraint}
\end{equation}
Taking the largest allowed value for $L\simeq 0.1$mm, 
the RS fine-tuning condition Eq.~(\ref{LLsLPl}) determines 
$L_s=(LL_{Pl}^2)^{1/3} \simeq 10^{-22}$mm
$\simeq 1/(10^6$TeV$)$ and $(L/L_s)^2 \simeq  10^{42}$
so that $y_s > L(L/L_s)^2 \simeq  10^{41}$mm $\sim  10^{16}$Mpc.
This is about 12 orders of magnitude larger than the present Hubble
scale. 
Also, since  $y_b(t)\ll L$ in the low energy regime, and
$y_s\gg L$ according to the inequality (\ref{e:constraint}), the
physical brane and the static brane need to be far apart at 
all times otherwise back reaction is not negligible.  
This situation is probably not very realistic. We need some high
energy, stringy effects to provoke the bounce and these may well
be relevant only when the branes are sufficiently close, i.e. at a
distance of order $L_s$. But in this case the constraint
(\ref{e:constraint}) will be violated which implies 
that back reaction will be relevant. 
On the other hand, if we want that $y_s \simeq L$ 
and back reaction to be unimportant, 
then Eq.~(\ref{e:constraint0}) implies that the  
bounce velocity has to be exceedingly small, $v_b \lsim 10^{-15}$.  
One might first hope to find a way out of these
conclusions by allowing the bounce to happen in the high 
energy regime.
But then $v_b \simeq 1$ and the nucleosynthesis bound is 
violated since too many zero-mode gravitons are being produced. 
Clearly our low energy approach looses its justification 
if $v_b \simeq 1$, but it seems unlikely that modifications 
coming from the high energy regime alleviate the bounds.     
\\
{\bf Conclusions:} 
Studying graviton production in an AdS braneworld we have 
found the following.
First, the energy density of KK gravitons
on the brane behaves as $\propto 1/a^6$, i.e. it scales like stiff
matter with the expansion of the Universe and can 
therefore not serve as a candidate for dark matter.
Furthermore, if gravity looks four dimensional on the brane, 
its higher-dimensional aspects, like the KK modes, 
are repelled from the brane. Even if KK gravitons are produced 
on the brane they rapidly escape into the bulk
as time evolves, leaving no traces of the underlying 
higher-dimensional nature of gravity.  
This is likely to survive also in other warped
braneworlds when expansion can be mimicked by 
brane motion.  
\\
Secondly, a braneworld bouncing at low energies
is not constrained by massless 4D gravitons and satisfies the
nucleosynthesis bound as long as $v_b \lsim 0.2$.   
However, for interesting values of the string and AdS scales
and the largest admitted bounce velocity 
the back reaction of the KK modes is only negligible 
if the two branes are far apart from each other at all times,
which seems rather unrealistic. 
For a realistic bounce the 
back reaction from KK modes can most likely not be neglected. 
Even if the energy density of the KK gravitons on the brane 
dilutes rapidly after the bounce, the corresponding energy 
density in the bulk could even lead to important changes of 
the bulk geometry. 
The present model seems to be adequate to address the back reaction
issue since the creation of KK gravitons happens 
exclusively at the bounce.    
This and the treatment of the high energy regime $v_b\simeq 1$ is 
reserved for future work.
\\
\\
We thank Kazuya Koyama for discussions.
This work is supported by the Swiss National Science Foundation.


\begin{thebibliography}{99}

\bibitem{Polchi}J. Polchinski, {\it String theory. An introduction to
  the bosonic string,  Vol. I}, and {\it String theory. Superstring
  theory and beyond, Vol. II }, Cambridge University Press (1998).\\
  J. Polchinski, Phys. Rev. Lett. {\bf 75}, 4724 (1995),
  \eprint{hep-th/9910219}.

\bibitem{RSI+II}L. Randall and R. Sundrum, Phys. Rev. Lett. {\bf 83},
  3370 (1999), \eprint{hep-th/9905221}; {\bf 83}, 4690, 
\eprint{hep-th/9906064}

\bibitem{bine}P. Binetruy, C. Deffayet, U. Ellwanger, and D. Langlois,
  Phys. Lett. {\bf B477}, 285 (2000),  \eprint{hep-th/9910219}.

\bibitem{casi} M. Ruser, Phys. Rev. {\bf A73}, 043811 (2006);
               J. Phys. {\bf A39}, 6711 (2006), and references therein. 

\bibitem{ekpy} J. Khoury, P. Steinhardt and N. Turok,
  Phys. Rev. Lett. {\bf 92}, 031302 (2004),
  \eprint{hep-th/0307132}; Phys. Rev. Lett. {\bf 91} 161301 (2003), 
\eprint{astro-ph/0302012}.

\bibitem{filo}R. Durrer and F. Vernizzi, Phys.Rev. {\bf D66}  083503 (2002),
  \eprint{hep-ph/0203275}; \
  C. Cartier, E. Copeland and R. Durrer, Phys. Rev.  {\bf D67},
  103517  (2003), \eprint{hep-th/0301198}.

\bibitem{roy}R. Maartens,
  Living Rev. Rel. {\bf 7}, 7 (2004), \eprint{gr-qc/0312059}.

\bibitem{myrev}R. Durrer, {\it Braneworlds},  at the XI
  Brazilian School of Cosmology and Gravitation, Edt. M. Novello and
  S.E. Perez Bergliaffa, AIP Conference Proceedings, 782 (2005),
  \eprint{hep-th/0507006}.

\bibitem{long} M.~Ruser and R.~Durrer, 
  Phys. Rev. D {\bf 76}, 104014 (2007), \eprint{arXiv:0704.0790}.

\bibitem{lang}M. Minamitsuji, M. Sasaki and D. Langlois, 
  Phys. Rev. {\bf D71}, 084019 (2005).

\bibitem{cyril}C. Cartier, R. Durrer and M. Ruser, 
  Phys. Rev. {\bf D72}, 104018 (2005).

\bibitem{michele}M. Maggiore. Phys. Rept. {\bf 331}, 283 (2000).

\end{thebibliography}
\end{document}